\documentclass[10pt,twocolumn,aps,prb,citeautoscript]{revtex4-1}
\usepackage[usenames,dvipsnames]{color}
\usepackage[pdftex]{graphicx}
\usepackage{amsmath}
\usepackage{amssymb}
\usepackage{bm}

\DeclareMathOperator{\tr}{Tr}

\renewcommand{\figurename}{\scriptsize {\bf Fig.}}
\renewcommand{\thefigure}{{\bf\arabic{figure}}}

\begin{document}

\title{\bf Sculpting stable structures in pure liquids}

\author{Tadej Emer\v{s}i\v{c}$^{1\ast}$}
\author{Rui Zhang$^{2\ast}$}
\author{\v{Z}iga Kos$^{3\ast}$}
\author{Simon \v{C}opar$^{3\ast}$}
\author{Natan Osterman$^{3,4}$}
\author{Juan J. de Pablo$^{2,5\dagger}$}
\author{Uro\v{s} Tkalec$^{1,4,6\dagger}$}

\affiliation{\footnotesize 
${}^1$ Institute of Biophysics, Faculty of Medicine, University of Ljubljana, Vrazov trg 2, 1000 Ljubljana, Slovenia.\\
${}^2$ Institute for Molecular Engineering, University of Chicago, Chicago, IL 60637, USA.\\
${}^3$ Faculty of Mathematics and Physics, University of Ljubljana, Jadranska 19, 1000 Ljubljana, Slovenia.\\
${}^4$ Jo\v{z}ef Stefan Institute, Jamova 39, 1000 Ljubljana, Slovenia.\\
${}^5$ Material Science Division, Argonne National Laboratory, Lemont, IL 60439, USA.\\
${}^6$ Faculty of Natural Sciences and Mathematics, University of Maribor, Koro\v{s}ka 160, 2000 Maribor, Slovenia.~{\rm $^\ast$These authors contributed equally to this work.\\ 
$^\dagger$Corresponding authors. E-mail: depablo@uchicago.edu (J.J.dP.); uros.tkalec@mf.uni-lj.si (U.T.)}
}

\begin{abstract}
\noindent Pure liquids in thermodynamic equilibrium are structurally homogeneous. In liquid crystals, flow and light pulses are used to create reconfigurable domains with polar order. Moreover, through careful engineering of concerted microfluidic flows and localized opto-thermal fields, it is possible to achieve complete control over the nucleation, growth, and shape of such domains. Experiments, theory, and simulations indicate that the resulting structures can be stabilized indefinitely, provided the liquids are maintained in a controlled non-equilibrium state. The resulting sculpted liquids could find applications in microfluidic devices for selective encapsulation of solutes and particles into optically active compartments that interact with external stimuli.
\end{abstract}

\maketitle
\begin{small}
\noindent Solid materials can simultaneously exhibit distinct structural phases, which can be manipulated to engineer functionality~\cite{Chaikin,Fert:NRevMat:17,Hellman:RMP:17}. Such structural phases, and the corresponding grain boundaries and defects, do not arise in pure liquids at equilibrium~\cite{Onuki}. On the one hand, liquids exhibit a number of attractive features, including short relaxation times, high diffusion coefficients, absolute compliance and, of course, the ability to wet surfaces~\cite{Debenedetti}. On the other hand, however, imbuing pure liquids with additional functionality is challenging due to their inherent homogeneity. Complex behaviour is generally encountered in multicomponent mixtures, be they synthetic or biological. In the particular case of biological systems, examples of self-organized, transient, and reconfigurable assemblies include raft domains, droplets and other membraneless compartments~\cite{Berry:RPP:18}. Such structures, however, are difficult to manipulate because they occur in out-of-equilibrium situations, and they generally involve multiple components that exhibit sharp miscibility gradients, leading to hydrophilic and hydrophobic domains. Recent efforts to ``print'' hydrophobic and hydrophilic domains into liquid mixtures by relying on the use of surfactant nanoparticles provide a striking demonstration of the possibilities afforded by gaining control over structure in non-equilibrium systems~\cite{Cui:Science:13}. Similarly, active matter with intrinsic topological defects is of particular relevance in this regard, where emergent structures can be created in the form of living colonies, tissues, and their biologically-inspired synthetic counterparts~\cite{Sanchez:Nature:12,Marchetti:RMP:13,Peng:Science:16,Huber:Science:18,Kumar:SciAdv:18}. In such examples, intrinsic activity leads to motion and transitions between different rheological regimes.

Recent advances in the ability to control the competing effects of confinement and external fields by purposely designed micromanipulation tools have enabled seminal studies of nucleation, stability and motion of topologically-protected configurations in complex fluids~\cite{Link:Science:97,Loudet:Nature:00,Iwashita:NMat:06,Gibaud:Nature:12,Nych:NPhys:17}, which can replace the need for multicomponent mixtures by creating distinct structural domains within a pure liquid. Liquid crystals (LCs) represent ideal systems for study of spontaneous symmetry breaking, topological defects, orientational ordering and phase transitions induced by applied stimuli~\cite{deGennes}. Even the simplest nematic phase, where the average orientation of rod-like molecules is characterized by a nematic director ${\bf n}$, exhibits a wide range of switching mechanisms between uniformly aligned states~\cite{Pieranski}. Nematic LCs (NLCs) can nucleate point and line defects~\cite{Alexander:RMP:12} by rapid pressure or temperature quenches~\cite{Bowick:Science:94} and in the presence of colloidal inclusions~\cite{Poulin:Science:97,Guzman:PRL:03,Musevic:Science:06,Tkalec:Science:11}. Recent work has shown that line defects can serve as microreactors in which to conduct polymerization reactions~\cite{Wang:NMat:16}, offering intriguing prospects for future applications. Thermodynamic and anchoring transitions, textures, hydrodynamics and flow instabilities in nematic mesophases have all been studied over the past decades. However, little is known about the coexistence and stability of driven orientational phases, and the corresponding defects under geometric confinement. Recent results on nematic flows in microfluidic environments~\cite{Cuennet:NPhot:11,dePablo:JCP:11,Kim:NCommun:12,Sengupta:PRL:13,Zhang:JCP:16,Giomi:PNAS:17} have raised the possibility of tuning multistable defect patterns, and the transitions that arise between flow regimes, by controlling the shape of the channels, the anchoring conditions on channel walls, the temperature-dependent material properties and, most importantly, the shear rate induced by the pressure differential that drives the flow.

\medskip
\noindent{\small\bf Results}\\
\noindent We demonstrate the creation and dynamic manipulation of defects and reconfigurable states in a pure NLC by simultaneous application of multiple external fields. Oriented polar phase domains are generated and controlled through combinations of confinement, flow, and laser pulses. In contrast to previous reports, where only bulk states of both homeotropic and flow-aligned phases were considered, we are able to observe the critical behavior of the phase interface for the first time. Precision control over the production, growth rate, and shape of the domains is achieved through a combination of simulations and experiments that involve flow control and opto-thermal manipulation driven by laser tweezers. More specifically, a quantitative analytical model of the phase interface is proposed to predict the behavior of both phases from the geometric and material parameters of the experimental setup. The model is supported by detailed numerical simulations that include all relevant structural and hydrodynamic details. The insights provided by our theoretical and computational work are used to design a responsive system in which a metastable flow-aligned phase is repeatedly reconfigured by switching the flow direction, thereby permitting detailed study of the nucleation of solitons, domain walls, and point defects, followed by their relaxation dynamics as they seek to return to equilibrium.

Distinct domains of a flow-aligned phase of pentylcyanobiphenyl (5CB) NLC in a linear microchannel are nucleated through a temperature quench with a laser beam. As shown in Fig.~1A, the channel has a rectangular cross-section and its walls confer perpendicular (homeotropic) surface alignment of the NLC (see Materials and Methods). The director profile of the initial stationary state corresponds to a uniformly aligned homeotropic configuration along the $z$ axis that appears black when observed between crossed polarizers. When the flow is turned on, the director remains predominantly aligned perpendicular to the substrate, but is slightly deflected in the flow direction due to flow alignment, changing the birefringent appearance from black to bright colors that depend on the flow velocity (Fig.~1B). We name this flow regime the ``bowser state'' , after the bowed shape of the director profile in contrast with the flow-aligned state, which we will refer to as the ``dowser state''. A black isotropic island is created where the NLC is heated into the isotropic phase by the laser tweezers; after the light is switched off and the NLC is quenched into the nematic phase, the initial tangle of defects relaxes into a flow-aligned state, bounded by a disclination loop (Fig.~1B). The flow-aligned domains then evolve with flow (Fig.~1C), and can either grow or annihilate depending on the flow velocity (Movies~S1 and S2). The flow-oriented phase, whose director field in the midplane of the channel is aligned horizontally, changes direction by a half-turn from the bottom to the top of the channel. This system represents an ideal experimental model of a quasi-two dimensional (2D) orientational phase, described by a unit vector field of the midplane director alignment, as recently reported in the strong flow regime $(v > 100~{\rm \mu m/s})$ in nematodynamics~\cite{Sengupta:PRL:13}, and it is analogous to the so-called dowser field in nematostatics, after which we adopted the name~\cite{Pieranski:PRE:16}.

\begin{figure}[!htb]
\centering \includegraphics[width=\columnwidth]{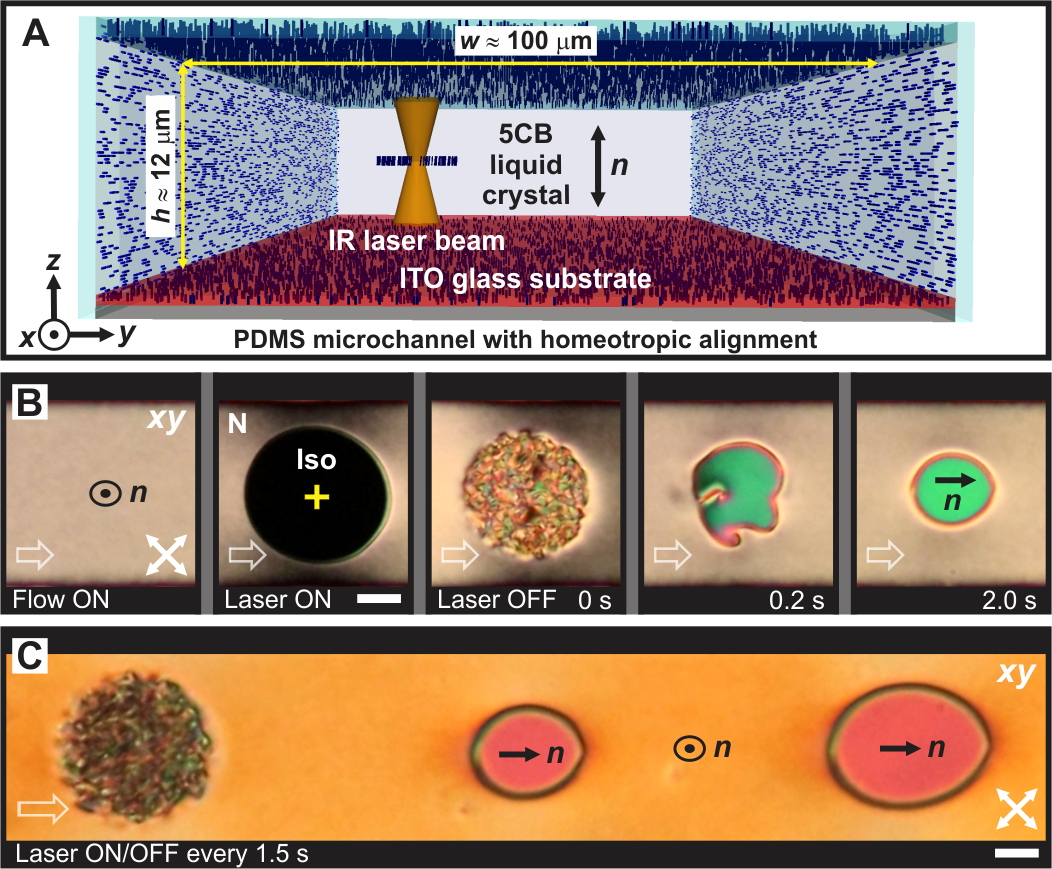}
\caption{
Nucleation of orientational phase domains in pressure-driven nematic microflows.
({\bf A}) Schematic illustration of a channel with homeotropic anchoring on the top and bottom surfaces, used in the experiment (see Materials and Methods).
({\bf B}) The nematic in a channel looks black between crossed polarizers in the absence of flow and gains visible birefringence due to flow-driven director distortion that traps a domain of the flow-aligned state (also called the ``dowser state'' from here on); {\bf n} denotes the nematic director. Strongly absorbed light of the laser tweezers heats the NLC, creating an isotropic (Iso) island which is quenched into the nematic (N) phase when the laser is switched off. The dense tangle of defects coarsen into a single defect loop that traps a flow-aligned dowser state, identifiable as a green area at low velocity.
({\bf C}) The laser-induced nucleation of dowser domains can be automated and their shape dynamically controlled by tuning the flow parameters. Crossed double arrows indicate orientation of the polarizers. White empty arrows in the bottom left corners indicate direction and qualitative velocity of the flow. Scale bars, $20~{\rm \mu m}$.
}
\label{Fig1}
\end{figure}

The dynamics of the director field, described by the in-plane angle $\phi$ relative to the $x$ axis, is driven by the elasticity and the flow alignment. Assuming uniform flow with midplane velocity $v$ along the $x$ axis and neglecting splay-bend anisotropy, we obtain a linear coupling of director and flow velocity

\begin{equation}
  \gamma_1\dot\phi=K\nabla^2\phi-\frac{2\left(\gamma_1-\gamma_2\right)}{\pi h}v\sin\phi \;.
  \label{eq:sine-gordon}
\end{equation}

\noindent This coupling is reminiscent of that of the dowser field to the thickness gradient~\cite{Pieranski:PRE:16,Pieranski:EPJE:16,Pieranski:EPJE:17}. The equation can be recognised as a damped sine-Gordon equation with characteristic length $\xi=\sqrt{\frac{K\pi h}{2v\left(\gamma_1-\gamma_2\right)}}$, where $K$ is single elastic constant,$\gamma_1$ and $\gamma_2$ are viscosity parameters, and where $h$ is the height of the channel. Details on the general derivation, including the velocity gradient term, which is essential for describing flow in channel junctions~\cite{Giomi:PNAS:17}, are provided in Materials and Methods.

The flow-aligned dowser state is stable under strong flows but unstable in weak flows. The dowser domains are seen to grow and shrink, depending on the flow velocity, as shown both in experiments (Fig.~2, A, C and F, and Movie~S2) and numerical simulations (Fig.~2, B, D and E, and Movie~S3). We have constructed a quantitative phenomenological model for the dynamics of the domain size (see Materials and Methods). In such a model, the growth rate of a circular domain with radius $r$ is given by

\begin{equation}
  \gamma_r\dot{r}=2\left(\gamma_1-\gamma_2\right)\left(v\cos\phi-v_c\right)-\frac{2\pi T}{r},
  \label{eq:rdot}
\end{equation}

\noindent where $T$ is the disclination line tension and $\gamma_r$ is the viscosity parameter associated with the drag force on moving disclination lines, which are both assumed to be constant. The critical velocity is $v_c=\frac{\pi^3K}{2h\left(\gamma_1-\gamma_2\right)}$. The stability of a flow-aligned ($\phi=0)$ dowser domain is conditioned by the midplane velocity in the channel, as summarized in the phase diagram of Fig.~2H. Below $v_c$, all dowser domains are unstable, and gradually shrink in size until they disappear. Above $v_c$, the behavior of dowser domains depends on their initial size. If the domain size is smaller than the critical radius $r_c=\frac{\pi T}{\left(\gamma_1-\gamma_2\right)}\left(v-v_c\right)^{-1}$, the domain shrinks over time and annihilates. If $r>r_c$, the domain grows until it reaches the side walls of the channels, and expands along the channel from then on (Fig.~2, A and B).

Equation~(\ref{eq:rdot}) can be integrated analytically to yield the time dependence of the flow-aligned dowser domain size

\begin{equation}
	t=\frac{1}{a}\left(r-r_0+r_c\ln\frac{r_c-r}{r_c-r_0}\right),
	\label{eq:relax}
\end{equation}

\noindent where $a=\frac{\gamma_1-\gamma_2}{\gamma_r}\left( v-v_c \right)$ and $r_0$ is the initial radius of the loop. We have fitted Eq.~(\ref{eq:relax}) to the experimental data in Fig.~2F and Fig.~S1 through parameters $a$, $r_0$ and $r_c$, and obtained good agreement with the model. From the fitting parameters, a $v_c$ of $(56.4 \pm 1.4)~{\rm \mu m/s}$ was determined as the point where the inverse of the $r_c$ reaches zero (Fig.~2G). Fitting over parameter $a$ yields a similar value $(56.8 \pm 1.2)~{\rm\mu m/s}$ (Fig.~S2). The critical velocity, calculated directly from the dimensions of a channel and the viscoelastic properties of 5CB, is $42~{\rm \mu m/s}$. The agreement with both values obtained from the fit is reasonable, particularly considering the simplifying assumptions of the theoretical model. A separate analysis, performed on channels of different sizes, reveals a comparable behavior and fits the model reasonably well (Fig.~S3).

\begin{figure*}[!htb]
\centering \includegraphics[width=2\columnwidth]{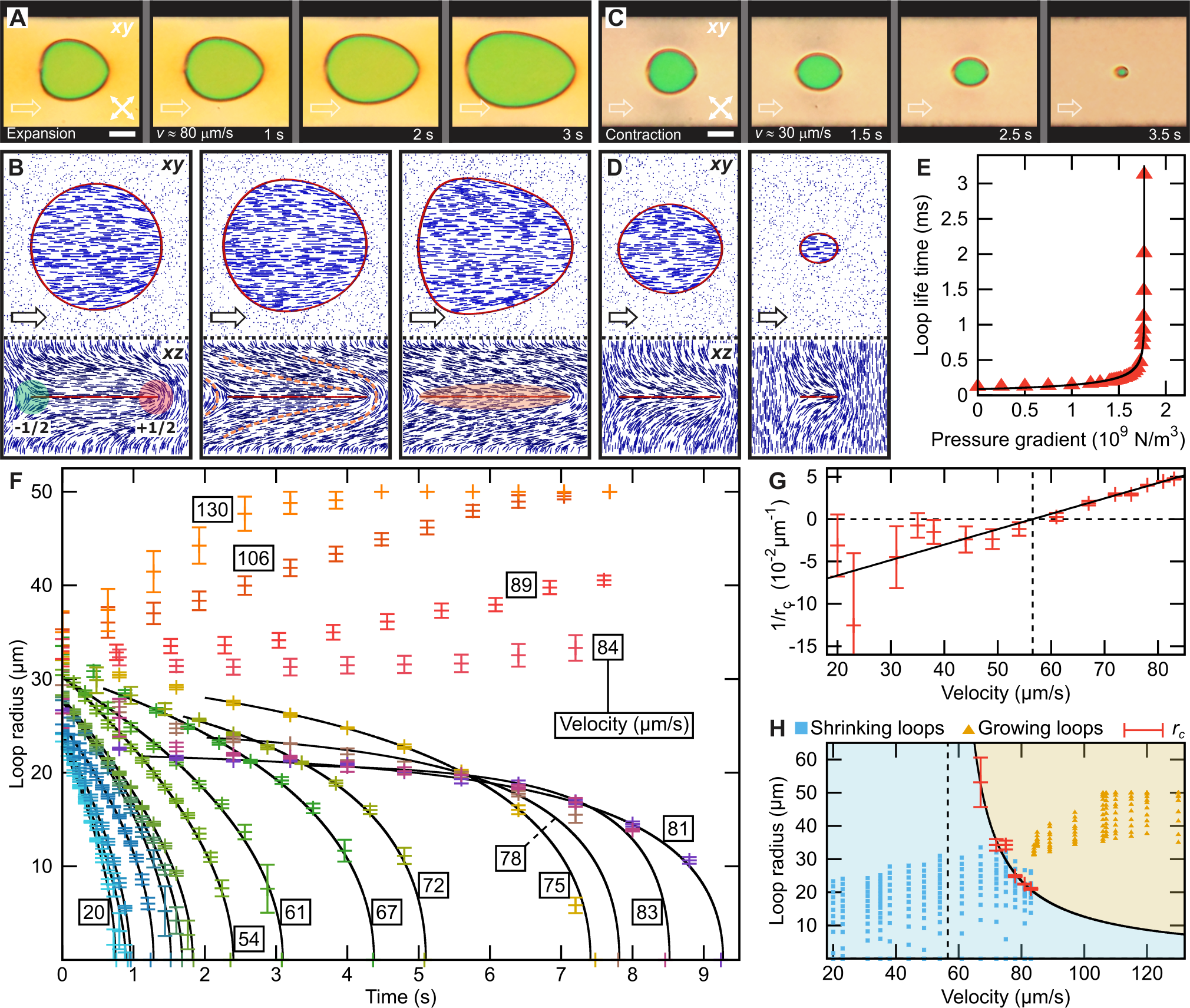}
\caption{
Dynamic evolution of dowser field domains in stationary nematic microflows.
({\bf A-D}) Growing and shrinking flow-aligned dowser domains in experiments and numerical simulations, captured at two different flow velocities. In panel (B), one can observe the varying profile of the half-integer disclination loop in the $xz$ plane, which serves as a phase boundary and stabilizes the dowser domain. Empty white arrows indicate the qualitative magnitude and direction of flow.
({\bf E}) Loop lifetime, determined from numerical simulations in the shrinking regime (D). The lifetime diverges at a certain critical pressure gradient that is proportional to the critical velocity. Note that the scale in simulations is orders of magnitude smaller than in experiments.
({\bf F}) Time dependence of the loop radius for different values of flow velocity. For shrinking loops, a theoretical model (Eq.~(\ref{eq:relax})) is fitted to the data points. The fitting function is shown by the bold lines. The theoretically predicted growth does not apply to growing loops, as their growth is confined by the channel walls.
({\bf G}) Critical velocity extracted from the fit parameter $r_{c}^{-1}$, obtained for loop annihilation at different velocities. A linear fit is used to determine the critical velocity at $(56.4 \pm 1.4)~{\rm\mu m/s}$.
({\bf H}) Phase diagram for shrinking (blue) and growing (orange) loops, separated by the curve for $r_c$ as obtained from the fit in panel (G). Some shrinking loop data points lie above the critical curve, due to loops that are still in the transition process after the quench and were thus omitted from the fit in panel (F). Scale bars, $20~{\rm \mu m}$.
}
\label{Fig2}
\end{figure*}

\noindent The dowser domains can be manipulated in various ways through careful application of the laser tweezers. A pre-existing bulk dowser state may be created upstream by prior quenching, or simply by the initial conditions of the nematic at the influx. One can produce a steady stream of domains by dissecting the original bulk dowser with a moving laser spot (Fig.~3A and Movie~S4) whose role is to constantly melt the sides of a phase boundary. A growing domain at higher flow velocity can be longitudinally split in half by a static laser beam at lower light intensities (Fig.~3B and Movie~S5). One can observe changing birefringent colors as the domain traverses a light-generated obstacle. The laser tweezers therefore enable dynamic control over the size, number, and lifetime of generated dowser domains, which can be further manipulated by periodic flow velocity modulations. In a uniform flow, the dowser field aligns uniformly along the flow direction and either grows or shrinks, depending on the velocity regime. By careful tuning and active control of the flow, a constant-size domain can be maintained over a period of tens of seconds (Fig.~3C). Here, the flowing dowser domain is stabilized in an oscillatory flow, maintained by a sinusoidal modulation of the driving pressure around the target flow rate ($70~{\rm \mu m/s} < v < 95~{\rm \mu m/s}$, with a period $3.5~{\rm s}$).

\begin{figure*}[!htb]
\centering \includegraphics[width=2\columnwidth]{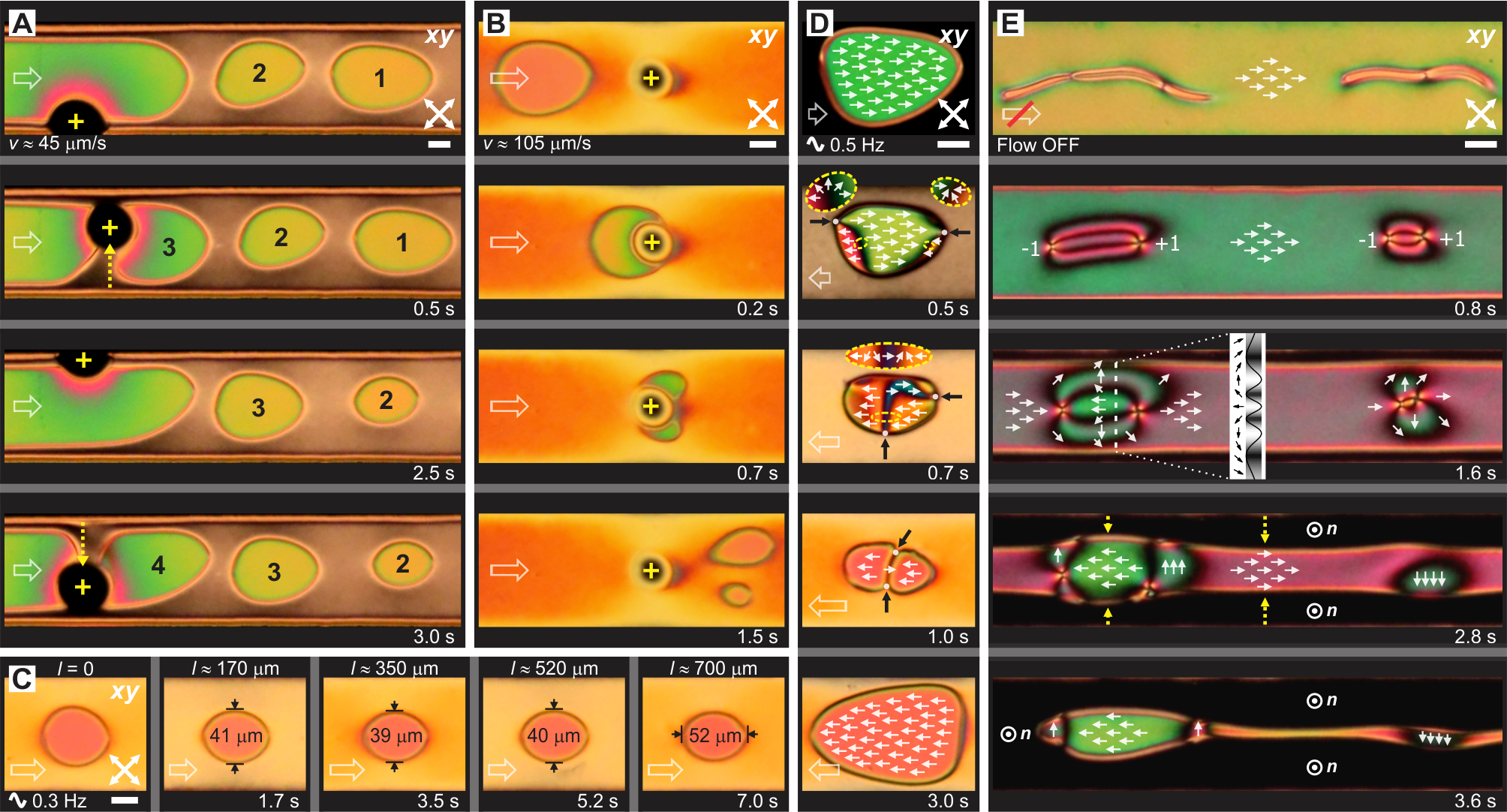}
\caption{
Systematic reshaping of dowser domains under laser action and oscillatory flows.
({\bf A}) Moving the laser beam transversely across the bulk dowser pinches off a uniform ``train'' of the domains.
({\bf B}) A static beam at a low power of $80~{\rm mW}$ generates a small isotropic region which cuts a large dowser domain longitudinally in half.
({\bf C}) The shape and size of the domain can be maintained over long time and length scales by periodically modulating the driving pressure around the value that induces the desired average flow rate.
({\bf D}) Under an alternating flow, a dowser domain reverses orientation every time the flow direction is changed. The reorientation creates surface point defects and realigning front, visible under the microscope as a rapid color change. The energetically unfavourable ``old'' orientation shrinks into a narrow $2\pi$ soliton and pinches the domain boundary (black arrows).
({\bf E}) Sufficiently rapid flow reversal creates point defect pairs connected by solitons. With the flow turned off, the characteristic length goes to infinity, and the solitons expand, revealing their signature profile in transmitted light intensity (inset). In a slow residual flow, flow-aligned parts shrink more slowly than parts with unfavourable orientation. Scale bars, $20~{\rm \mu m}$.
}
\label{Fig3}
\end{figure*}

Our model, as expressed in Eq.~(\ref{eq:sine-gordon}), predicts that, when the flow direction is reversed, the previous equilibrium state becomes the least favourable, leading to a rapid reversal of orientation. Figure~3D and Movie~S6 show a dowser domain in an alternating flow where flow reversal can be observed twice per period ($-90~{\rm \mu m/s} < v < +90~{\rm \mu m/s}$ with a period $2.0~{\rm s}$). When the velocity drops below the critical value, the domain begins to shrink. As the flow reverses, the reorientation begins at the leading and trailing points of the domain, creating reorientation fronts that propagate inwards (Fig.~3D). The part with the old orientation is energetically unfavourable, causing it to shrink to a narrow $2\pi$ soliton. When the velocity passes $v_c$, the domain starts to grow again, leading to disappearance of solitons through discontinuous director rearrangement (Fig.~S4). The soliton size is set by the characteristic length $\xi$ from the sine-Gordon equation, which becomes very small under strong flows.

The line tension Eq.~(\ref{eq:rdot}) holds locally and describes the curvature flow of the dowser domain boundary. When the dowser orientation $\phi$ is not aligned with the flow, the domain growth rate varies along the boundary and may even change sign. Figure~3D shows invaginations that appear in the dowser domain, where the soliton extends to the boundary. The boundary of the domain is thus controlled by the internal orientation, leading to non-circular shapes that are reminiscent of living cells during division. In Fig.~3E, one can appreciate that bulbous parts appear on the shrinking domains; these are due to small residual flows that cause a difference in line tension between regions with different director orientations.

The dowser state, being a polar unit vector field, supports the existence of point defects with an integer winding number, which appear in pairs, connected by a soliton. As shown in Fig.~3E and Movie~S7, turning off the flow increases the characteristic length with decreasing velocity, allowing us to clearly observe the detailed structure of the soliton. In this simple case, the sine-Gordon equation can be solved analytically, giving the transverse profile of the soliton as $\phi(y)=4\arctan e^{\pm y/\xi}$ (inset in Fig.~3E), analogous to the soliton profile seen in the dowser field under the cuneitropism effect~\cite{Pieranski:PRE:16}. In addition to defect pairs, closed circular solitons may appear during oscillatory flow (Fig.~S4).

\bigskip
\noindent{\small\bf Discussion}\\
\noindent In close analogy to other heterogeneous mixtures with phase boundary interfaces, such as fogs and aerosols, microfluidic droplets, fluids at the triple point, and the nematic-isotropic phase transition itself, surface tension plays a crucial role in controlling the ``evaporation'' and growth of the low energy phase, depending on the curvature of the interface. In the system presented here, the phases possess different symmetry, are topologically incompatible, and exhibit what is essentially a first-order phase transition. The dowser state is orientationally anisotropic, with its own elastic behavior, topological defects, and solitons, whereas the bowser state is effectively isotropic and simple in the simplified 2D view. The dowser phase orientation couples with the velocity field, and can have either lower or higher energy than the bowser, depending on its orientation. In this way, the shape, splitting, and coalescence of domains can be controlled. 

As the dowser field also couples to external magnetic and electric fields, as well as the gradients of the channel thickness, we suggest that combinations of such external cues could offer exceptional possibilities for shape control, flow steering and optical tuning. As the response to the external stimuli is directly and clearly observable through birefringence, it could also be used to measure viscoelastic and rheological properties of the material itself. The dowser domains, readily produced in large quantities and sizes with laser tweezers, are analogous to droplets and shells, and can serve similar purposes. It should be possible, for example, to create nested domains for production of 2D shells with an enclosed volume of the dowser phase. It should also be possible to conduct chemical reactions in such domains, or sequester different components according to orientational or dielectric affinity (as opposed to hydrophobic/hydrophilic contrast), as recently shown for defects around microparticles \cite{Wang:NMat:16}. One could envision a 3D printing system for liquids in which complex out-of-equilibrium structures are created and stabilized by relying on the principles outlined here. Finally, the models obtained here from experiments with standard thermotropic LCs could be readily applied to active and biological materials with nematic behavior. Controlling alignment and separation into isolated orientational domains may be seen as a technical tool with potential for applications in biophysics, chemistry, and chemical engineering.
\end{small}

\bigskip
\begin{scriptsize}
\noindent{\small\bf Materials and Methods}\\
\noindent{\bf Materials and experimental procedures.} We have used a single-component nematic material 5CB (SYNTHON Chemicals), which has the nematic phase in temperature range $18~{^\circ}{\rm C} < T < 35~{^\circ}{\rm C}$, in all of our experiments. The microfluidic channels had rectangular cross-section, with height $h \approx 12~{\rm \mu m}$, width $w = 100~{\rm \mu m}$, and length $L = 20~{\rm mm}$. The channels were fabricated out of polydimethylsiloxane (PDMS, Sylgard 184, Dow Corning) reliefs and indium-thin-oxide (ITO) coated glass substrates (Xinyan Technology) by following standard soft lithography procedures~\cite{Sengupta:PRL:13}. The ITO coating was used as an absorber of the infrared (IR) laser light and provided very good control for the local heating of 5CB. The channel walls were chemically treated with $0.2~{\rm wt\%}$ aqueous solution of silane dimethyl-octadecyl-3-aminopropyl-trimethoxysilyl chloride (DMOAP, ABCR) to induce strong homeotropic surface anchoring for 5CB molecules~\cite{Sengupta:PRL:13}. The microfluidic channels were filled with 5CB in its isotropic phase and allowed to cool down to nematic phase at room temperatute before starting the flow experiments. We drove and precisely controlled the fluid flow by using a pressure-driven microfluidic flow control system (OB1, Elveflow). We applied and varied flow rates in the range $[0.05, 1.50]~{\rm \mu L/h}$ corresponding to a flow velocity $v$, ranging $5 - 250~{\rm \mu m/s}$. In some experiments, the nematodynamics was controlled by adjusting time-dependent flow driving using in-built flow profile routines of the microfluidic controller. The characteristic Reynolds number $Re = \rho v l / \eta$ for 5CB having effective dynamic viscosity~\cite{Giomi:PNAS:17} $\eta \approx 50~{\rm mPas}$ ranged between $10^{-6}$ and $10^{-4}$. Here, $\rho \approx 1.024~{\rm kg/m^3}$ is the material's density, and $l = 4wd/2(w+h) \approx 21~{\rm \mu m}$ is the hydraulic diameter of the channels. The corresponding Ericksen number $Er = \eta v l/K$, with $K = 5.5~{\rm pN}$ being the 5CB single elastic constant approximation, varied between $0.8$ and $40$. All the experiments were conducted at room temperature.
\medskip

\noindent{\bf Polarized light microscopy, laser tweezers and image acquisition.} The flow regimes, the reorientation dynamics and the flow-driven deformations of 5CB in microchannels with homeotropic surface anchoring were studied by polarized light microscopy (Nikon, Eclipse Ti-U, equipped with CFI Plan $2\times$ and $10\times$ objectives). The samples were observed between crossed polarisers in transmission mode. In addition, we used a laser tweezers setup build around the inverted optical microscope with an IR fiber laser operating at $1064~{\rm nm}$ as a light source and a pair of acousto-optic deflectors driven by computerized system (Aresis, Tweez 200si) for precise laser beam manipulation. The laser power was varied between $20~{\rm mW}$ and $200~{\rm mW}$ in the sample plane and the Gaussian beam profile was primarily used for local heating of the NLC above its clearing temperature. Full HD color videos were recorded at a frame rate of $30$ frames per second using a digital CMOS camera (Canon, EOS 750D), attached by a C-mount compatible adapter (LMScope) to the microscope. The image analysis was performed using the software ImageJ.
\medskip

\noindent{\bf Numerical simulation details.} The bulk free energy of the NLC, $F$, is defined as

\begin{equation}
\begin{aligned}
F&= \int_{V} f_\text{bulk}\,\mathrm{d}V + \int_{\partial V} f_\text{surf} \,\mathrm{d}S\\
&= \int_{V} \left( f_\text{LdG}+f_\text{el} + f_\text{D} \right) \mathrm{d}V + \int_{\partial V} f_\text{surf} \,\mathrm{d}S,
\end{aligned}
\label{total}
\end{equation}

\noindent where $f_\text{LdG}$ is the short-range free energy, $f_\text{el}$ is the long-range elastic energy, $f_\text{D}$ is laser-induced dielectric interaction energy, and $f_\text{surf}$ is the surface free energy due to anchoring. $f_\text{LdG}$ is given by a Landau-de Gennes expression of the form~\cite{deGennes}:

\begin{equation}
f_\text{LdG}= \frac{A_0}{2} \left( 1-\frac{w}{3} \right) \tr({\bf Q}^2) - \frac{A_0 w}{3} \tr({\bf Q}^3) + \frac{A_0 w}{4}(\tr({\bf Q}^2))^2.
\label{phase}
\end{equation}

\noindent Parameter $w$ controls the magnitude of $q_0$, namely the equilibrium scalar order parameter via $q_0=\frac{1}{4}+\frac{3}{4}\sqrt{1-\frac{8}{3w}}$. The elastic energy $f_\text{el}$ is written as 
(using Einstein summation rule):

\begin{equation}
\begin{aligned}
f_\text{el}=&\frac{1}{2}L_1 (\partial_kQ_{ij})(\partial_kQ_{ij})+\frac{1}{2}L_2 (\partial_kQ_{jk})(\partial_lQ_{jl})\\
&+\frac{1}{2}L_3 Q_{ij}(\partial_iQ_{kl})(\partial_jQ_{kl})+\frac{1}{2}L_4 (\partial_lQ_{ik})(\partial_kQ_{jl}),
\end{aligned}
\label{elastic_en}
\end{equation}

\noindent where $\partial_i$ indicates a spatial derivative over $i$-th coordinate.
If the system is uniaxial, the $L$'s in Eq.~(\ref{elastic_en}) can be mapped to the Frank elastic constants $K$'s via

\begin{equation}
\begin{aligned}
L_1&=\frac{1}{2q_0^2} \left[ K_{22}+\frac{1}{3}(K_{33}-K_{11}) \right], \\
L_2&=\frac{1}{q_0^2} (K_{11}-K_{24}), \\
L_3&=\frac{1}{2q_0^3} (K_{33}-K_{11}), \\
L_4&=\frac{1}{q_0^2} (K_{24}-K_{22}).
\end{aligned}
\end{equation}

\noindent By assuming a single elastic constant $K_{11}=K_{22}=K_{33}=K_{24}\equiv K$, one has $L_{1}=L\equiv K/2q_0^2$ and $L_{2}=L_{3}=L_{4}=0$. Pointwise, ${\bf n}$ is the eigenvector associated with the greatest eigenvalue of the ${\bf Q}$-tensor at each lattice point. The free energy associated with anisotropic dielectric constants reads

$$ f_\text{D}=-\frac{1}{2} \epsilon_0 \epsilon_{ij} E_i E_j \mathrm{d}V, $$

\noindent where $\epsilon_0$ is the vacuum permittivity constant, and $\epsilon_{ij}$ is the dielectric permittivity tensor related to the {\bf Q}-tensor as

$$ \epsilon_{ij} = \bar\epsilon \, \delta_{ij} + \frac{2}{3} (\epsilon_{||}-\epsilon_{\perp})Q_{ij}, $$

\noindent in which $\epsilon_{||}$ and $\epsilon_{\perp}$ are the permittivities parallel and perpendicular to the nematic director, respectively.

To simulate NLC's flowing dynamics, a hybrid lattice Boltzmann method is used to simultaneously solve a Beris-Edwards equation and a momentum equation which accounts for the back-flow effects. By introducing a velocity gradient $W_{ij}=\partial_i v_j$, strain rate ${\bf u}=\frac{1}{2}({\bf W} + {\bf W}^T)$, vorticity tensor $\boldsymbol{\omega}=\frac{1}{2}({\bf W} - {\bf W}^T)$, and a generalized advection term

\begin{equation}
\begin{aligned}
{\bf S}({\bf W},{\bf Q})=&(\xi {\bf u}-\boldsymbol\omega)({\bf Q}+{\bf I}/3)+({\bf Q}+{\bf I}/3)(\xi {\bf u}+\boldsymbol\omega)&\\
&-2\xi ({\bf Q}+{\bf I}/3) \tr({\bf QW}),&
\end{aligned}
\end{equation}

\noindent one can write the Beris-Edwards equation~\cite{Beris} according to

\begin{equation}
(\partial_t +{\bf v}\cdot \nabla){\bf Q}-{\bf S}({\bf W},{\bf Q})=\Gamma \bf{H},
\label{beris_edwards_eq}
\end{equation}

\noindent where $\partial_t$ is a partial derivative over time. The constant $\xi$ is related to the material's aspect ratio and relates to the alignment parameter in the Ericksen-Leslie-Parodi theory $\lambda=-\frac{\gamma_2}{\gamma_1}=(2+q_0)\xi/3q_0$. $\Gamma$ is related to the rotational viscosity $\gamma_1$ of the system by $\Gamma=2q_0^2/\gamma_1$~\cite{Denniston:PRE:01}. The molecular field $\bf{H}$, which drives the system towards thermodynamic equilibrium, is given by

\begin{equation}
{\bf H}=-\left[ \frac{\delta F}{ \delta \bf{Q}} \right]^\text{st},
\end{equation}

\noindent where $\left[ ...\right]^\text{st}$ is a symmetric and traceless operator. When velocity is absent, i.e. ${\bf v}({\bf r})\equiv 0$, Beris-Edwards equation Eq.~(\ref{beris_edwards_eq}) reduces to Ginzburg-Landau equation:

$$ \partial_t {\bf Q}=\Gamma {\bf H}. $$

\noindent To calculate the static structures of $\pm 1/2$ defects, we adopt the above equation to solve for the ${\bf Q}$-tensor at equilibrium.

Homeotropic anchoring is implemented through a Rapini-Papoular expression~\cite{deGennes} that penalizes any deviation of the ${\bf Q}$ tensor from ${\bf Q}_0$, namely a surface-preferred {\bf Q}-tensor. The associated free energy expression is given by

\begin{equation}
f_\text{surf} = \frac{1}{2}W ({\bf Q} - {\bf Q}_0)^2.
\end{equation}

\noindent The evolution of the surface ${\bf Q}$-field is governed by~\cite{Zhang:JCP:16}:

\begin{equation}
\frac{\partial {\bf Q}}{\partial t}=-\Gamma_s \left( -L {\boldsymbol \nu} \cdot \nabla {\bf Q} +\left[  \frac{\partial f_\text{surf}}{\partial {\bf Q}} \right]^\text{st} \right),
\label{surface_evolution}
\end{equation}

\noindent where $\Gamma_s=\Gamma/\xi_N$ with $\xi_N=\sqrt{L_1/A_0}$, namely nematic coherence length. The above equation is equivalent to the mixed boundary condition given in Ref.~\cite{Batista:SM:15} for steady flows.

The momentum equation for the nematics is written as~\cite{Denniston:PRE:01}

\begin{equation}  
\begin{aligned}
\rho(\partial_t+v_j \partial_j)v_i=&\partial_j \Pi_{ij}+\eta\partial_j[\partial_i v_j+\partial_j v_i\\
&+(1-3\partial_\rho P_0)\partial_k v_k \delta_{ij}].
\end{aligned}
\label{ns_eq}
\end{equation}

\noindent The stress ${\bf \Pi}$ is defined as

\begin{equation}
\begin{aligned}
\Pi_{i j}=  & -P_0 \delta_{i j}-\xi H_{i k} \left( Q_{k j} +\frac{1}{3}\delta_{k j} \right) - \xi \left( Q_{i k} +\frac{1}{3} \delta_{i k} \right) H_{k j} & \\
 &+ 2 \xi \left( Q_{i j} +\frac{1}{3}\delta_{i j} \right) Q_{k l}H_{k l} -\partial_{j} Q_{k l} \frac{\delta \mathcal F}{\delta \partial_i Q_{k l}} &\\
 & + Q_{i k} H_{k j} -H_{i k} Q_{k j}, &
\end{aligned}
\label{stress}
\end{equation}

\noindent where $\eta$ is the isotropic viscosity, and the hydrostatic pressure $P_0$ is given by

\begin{equation}
P_0=\rho T - f _\text{bulk}.
\end{equation}

\noindent The temperature $T$ is related to the speed of sound $c_s$ by $T=c_s^2$. We solve the evolution equation Eq.~(\ref{beris_edwards_eq}) using a finite-difference method. The momentum equation Eq.~(\ref{ns_eq}) is solved simultaneously via a lattice Boltzmann method over a D3Q15 grid~\cite{Guo}. The implementation of stress follows the approach proposed by Guo {\sl et al.}~\cite{Guo:PRE:02}. Our model and implementation were validated by comparing our simulation results to predictions using the Ericksen-Leslie-Parodi theory~\cite{Leslie:ALC:79}. The units are chosen as follows: the unit length $a$ is chosen such that unit length $a=\xi_N=7~{\rm nm}$, viscosity $\gamma_1=0.07~{\rm Pa s}$, and elastic constants: splay $K_{11}=6~{\rm pN}$, twist $K_{22}=3.9~{\rm pN}$, bend $K_{33}=8.2~{\rm pN}$, and saddle-splay $K_{24}=7~{\rm pN}$, mimicking the material properties of 5CB. We refer the reader to Ref.~\cite{Zhang:JCP:16} for additional details on the numerical methods employed here.
\medskip

\noindent{\bf Dowser field orientation.} To derive an effective 2D theory of a dowser state, we take the ansatz for the dowser profile of the director field

\begin{equation}
{\bf n}={\bf d}\sin\left(\frac{z\pi}{h}\right)+{\bf e}_z\cos\left(\frac{z\pi}{h}\right),
\label{eq:Dowser_ansatz}
\end{equation}

\noindent where unit vector ${\bf d}$ is a dowser orientational field in the $xy$ plane, $z$ the vertical position in the channel of height $h$, and ${\bf e}_z$ a unit vector along $z$ direction. A Poiseuille flow profile is assumed

\begin{equation}
{\bf v}(z)=4\left(1-\frac{z}{h}\right)\frac{z}{h}{\bf v},
\label{eq:Poiseuille_ansatz}
\end{equation}

\noindent where ${\bf v}$ is a vector field in the $xy$ plane of the channel. Given that we are interested only in the behaviour of the director field, we write the dissipation function, omitting the terms that do not include time derivatives of ${\bf n}$~\cite{Vertogen}

\begin{equation}
D=\alpha_2\dot{n}_in_j\partial_jv_i+\alpha_3\dot{n}_in_j\partial_iv_j+\frac{1}{2}\gamma_1\dot{n}_i\dot{n_i},
\end{equation}

\noindent where $\alpha_2=\frac{1}{2}\left(\gamma_2-\gamma_1\right)$ and $\alpha_3=\frac{1}{2}\left(\gamma_2+\gamma_1\right)$ are viscosity coefficients in the Ericksen-Leslie-Parodi formulation of nematodynamics and a dot indicates the time derivative. Total dissipation in the system is given as a volume integral over the dissipation function $D$. Since the director and the velocity profile in the $z$ direction are fixed by Eq.~(\ref{eq:Dowser_ansatz}) and Eq.~(\ref{eq:Poiseuille_ansatz}), respectively, we are free to perform the integration over the $z$ axis, obtaining effectively a  dissipation function $D_{\rm 2D}=\int D\mathrm{d}z$ of 2D processes

\begin{equation}
\begin{split}
D_{\rm 2D}=&\alpha_2h\dot{d}_id_j\partial_jv_i\left(\frac{1}{3}+\frac{1}{\pi^2}\right)+\alpha_2\dot{d}_iv_i\frac{2}{\pi}\\
&+\alpha_3h\dot{d}_id_j\partial_iv_j\left(\frac{1}{3}+\frac{1}{\pi^2}\right)+\frac{\gamma_1}{4}h\dot{d}_i\dot{d}_i.
\end{split}
\end{equation}

\noindent Next, we write the elastic free energy density of a dowser structure in a one elastic constant ($K$) approximation $f_\text{el}=\frac{K}{2}\left[\left(\nabla\cdot{\bf n}\right)^2+\left(\nabla\times{\bf n}\right)^2\right]$, which can again be integrated over the $z$ axis $f_{\rm 2D}=\int f_\text{el}\mathrm{d}z$, obtaining

\begin{equation}
f_{\rm 2D}=\frac{K}{2}\left[\frac{h}{2}\left(\nabla\cdot{\bf d}\right)^2-\pi\left(\nabla\cdot{\bf d}\right)+\frac{h}{2}\left(\nabla\times{\bf d}\right)^2+\frac{\pi^2}{h}\right].
\end{equation}

\noindent We write dowser field as ${\bf d}=(\cos\phi,\sin\phi)$ and follow the Lagrange formalism for $\phi$ angle $\partial_i\frac{\partial f_{\rm 2D}}{\partial(\partial_i\phi)}-\frac{\partial f_{2D}}{\partial \phi}=\frac{\partial D_{\rm 2D}}{\partial\dot{\phi}}$, obtaining the master equation for the dowser orientation

\begin{equation}
\begin{split}
\dot{\phi}&=\frac{K}{\gamma_1}\nabla^2\phi+\frac{K\pi}{\gamma_1h}\left({\bf d}\cdot\nabla\right)\phi+2\left(\frac{1}{3}+\frac{1}{\pi^2}\right)\omega_{xy}\\
&\phantom{{}=}-2\lambda\left(\frac{1}{3}+\frac{1}{\pi^2}\right)\left(u_{xx}\sin 2\phi-u_{xy}\cos 2\phi\right)\\
&\phantom{{}=}-(\lambda+1)\frac{2}{\pi h}\left(v_x\sin\phi-v_y\cos\phi\right).
\end{split}
\end{equation}
\medskip

\noindent{\bf Effective free energy of dowser domains.} Time derivative of the dowser orientation can be viewed also as a relaxation under the effective potential $U$ that depends on the velocity field. The equation of motion in that case is

\begin{equation}
\frac{\gamma_1}{2}h\dot{\phi}-h\gamma_1\left(\frac{1}{3}+\frac{1}{\pi^2}\right)\omega_{xy}=\partial_i\frac{\partial f_{\rm 2D}}{\partial(\partial_i\phi)}-\frac{\partial U}{\partial\phi}.
\end{equation}

\noindent The master equation for $\dot{\phi}$ is recovered for

\begin{equation}
\begin{split}
\frac{\partial U}{\partial\phi}&=-\gamma_2h\left(\frac{1}{3}+\frac{1}{\pi^2}\right)\left(u_{xx}\sin 2\phi-u_{xy}\cos 2\phi\right)\\
&\phantom{{}=}-(\gamma_2-\gamma_1)\frac{1}{\pi}\left(v_x\sin\phi-v_y\cos\phi\right).
\end{split}
\end{equation}

\noindent Integrating the above equation over $\phi$ leads to

\begin{equation}
\begin{split}
U&=\frac{\gamma_2h}{2}\left(\frac{1}{3}+\frac{1}{\pi^2}\right)\left(u_{xx}\cos 2\phi+u_{xy}\sin 2\phi\right)\\
&\phantom{{}=}+\frac{(\gamma_2-\gamma_1)}{\pi}\left(v_x\cos\phi+v_y\sin\phi\right)+C,
\end{split}
\end{equation}

\noindent which can be written in a covariant form

\begin{equation}
U=\frac{\gamma_2h}{2}\left(\frac{1}{3}+\frac{1}{\pi^2}\right){\bf d} {\sf u} {\bf d}^\intercal
+\frac{(\gamma_2-\gamma_1)}{\pi}\,{\bf v}\cdot{\bf d}+C,
\end{equation}

\noindent where a constant $C$ is not dependent on $\phi$. $C$ has to vanish at ${\bf v}=0$ and must be at least quadratic in $v$ to preserve the invariant form. Since we are interested only in the linear response to the velocity field, we can set $C=0$. 

Using the effective potential $U$, we can phenomenologically construct an effective free energy $F$ of a dowser state in microfluidic confinement in contact with a homeotropic nematic state (with ansatz ${\bf n}={\bf e}_z$):

\begin{equation}
F=\int\displaylimits_{\substack{\text{dowser}\\ \text{area}}}\left(f_{\rm 2D}+U\right)\mathrm{d}S
+\int\displaylimits_{\substack{\text{dowser}\\ \text{edge}}}T\mathrm{d}l,
\end{equation}

\noindent where $T$ is line tension of a nematic disclination~\cite{deGennes}. Specifically, we are interested in the free energy of circular dowser domains with homogeneous alignment of the dowser vector ${\bf d}$ in flow field that is homogenous in the $xy$ plane. This substantially simplifies the expression for the free energy

\begin{equation}
F=-\left(\gamma_1-\gamma_2\right)\left({\bf v}\cdot{\bf d}-v_c\right)r^2+2\pi Tr,
\end{equation}

\noindent where $v_c=\frac{\pi^3K}{2h\left(\gamma_1-\gamma_2\right)}$. The dynamics of the loop growth or annihilation is given by

\begin{equation}
\gamma_r\dot{r}=-\frac{1}{r}\frac{\partial F}{\partial r},
\end{equation}

\noindent where viscosity parameter $\gamma_r$ is due to a drag force on a moving disclination line.

The line tension contribution becomes dominant at small radii, leading to universal annihilation behaviour of shrinking loops (Fig.~S1).

\noindent{\small\bf Acknowledgements}\\
\noindent We thank J. A. Martinez-Gonzalez and M. Ravnik for stimulating discussions on topics related to this work. {\bf Funding:} This work was conducted with support from the Slovenian Research Agency (ARRS) under contracts P1-0055 (to T.E. and U.T.), P1-0099 (to \v{Z}.K. and S.\v{C}.), P1-0192 (to N.O.), J1-6723 (to U.T.), L1-8135 (to \v{Z}.K. and S.\v{C}.), and J1-9149 (to S.\v{C}.), and with support from the National Science Foundation of the U.S.A. under grant DMR 1710318 (to R.Z. and J.J.dP.). U.T. would like to thank COST Action MP1205 ``Advances in Optofluidics: Integration of Optical Control and Photonics with Microfluidics'' and COST Action MP1305 ``Flowing Matter'' for supporting his research activities.  {\bf Author contributions:} U.T. and J.J.dP. designed the research. T.E. and U.T. conducted the experiments, R.Z. and J.J.dP. performed the numerical simulations and analysed the data, \v{Z}.K. and S.\v{C}. developed theoretical model and analysed the results. N.O. contributed measurements for the Supplementary Materials. U.T., S.\v{C}., \v{Z}.K., R.Z., and J.J.dP. wrote the paper; U.T. and J.J.dP. supervised the research. All authors discussed the progress of research and contributed to the final version of the manuscript. Additional data related to this paper may be requested from the authors.\vfill
\end{scriptsize}

\onecolumngrid
\pagestyle{empty}
\renewcommand{\figurename}{Fig.}
\renewcommand{\thefigure}{S\arabic{figure}} 

\begin{small}
\section*{Supplementary Figures}

\setcounter{figure}{0}
\begin{figure*}[!htb]
\centering \includegraphics[width=105mm]{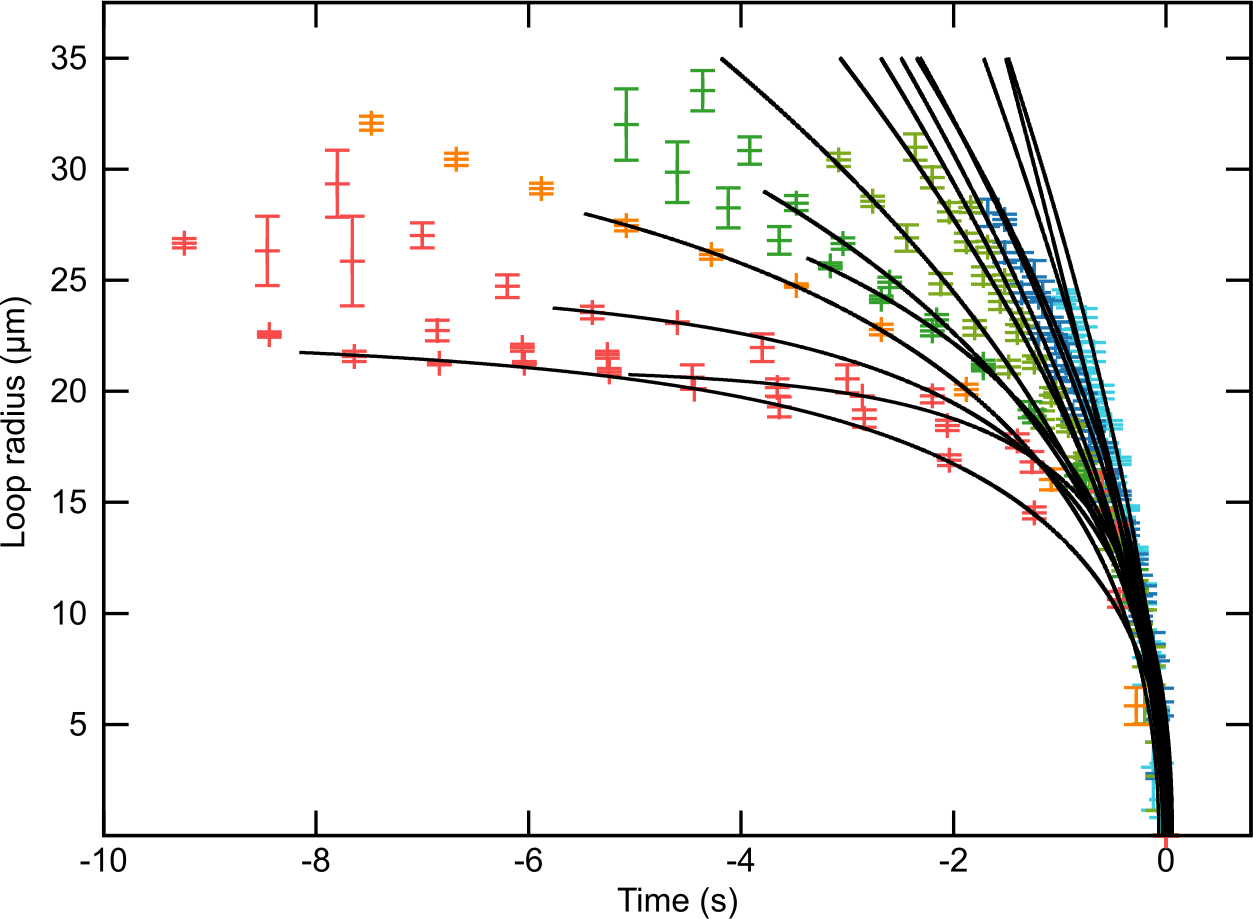}
\caption{\small
{\bf Universality of the dowser domain annihilation dynamics at small radii.} Experimental data points and the corresponding theoretical curves for the shrinking defect loop dynamics in Fig.~2F can be time-shifted to align the annihilation times, showing universal behaviour at small radii, when loop dynamics is governed primarily by the elasticity effects.
}
\end{figure*}

\begin{figure}[!htb]
\centering \includegraphics[width=85mm]{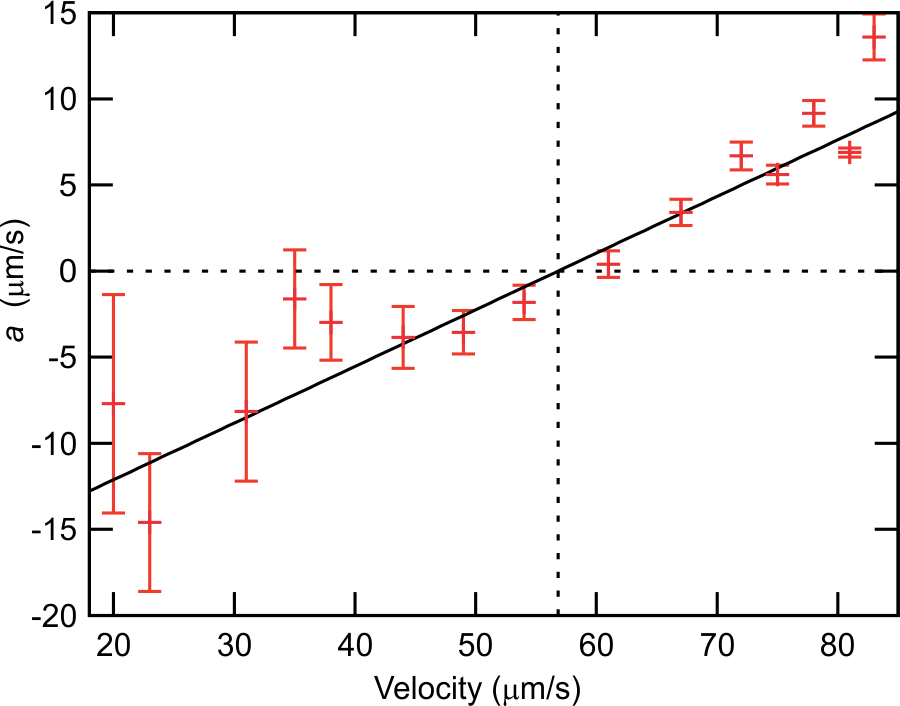}
\caption{\small
{\bf Velocity dependence of parameter $a$ from Eq.~(3) for the dowser domain size, as extracted from Fig.~2F.} A critical velocity of $v_c = (56.8 \pm 1.2)\,\mu\text{m/s}$ is extracted from the linear fit, which corresponds well to the value of $v_c$ obtained in Fig.~2G.
}
\end{figure}
\clearpage

\begin{figure*}[!htb]
\centering \includegraphics[width=152mm]{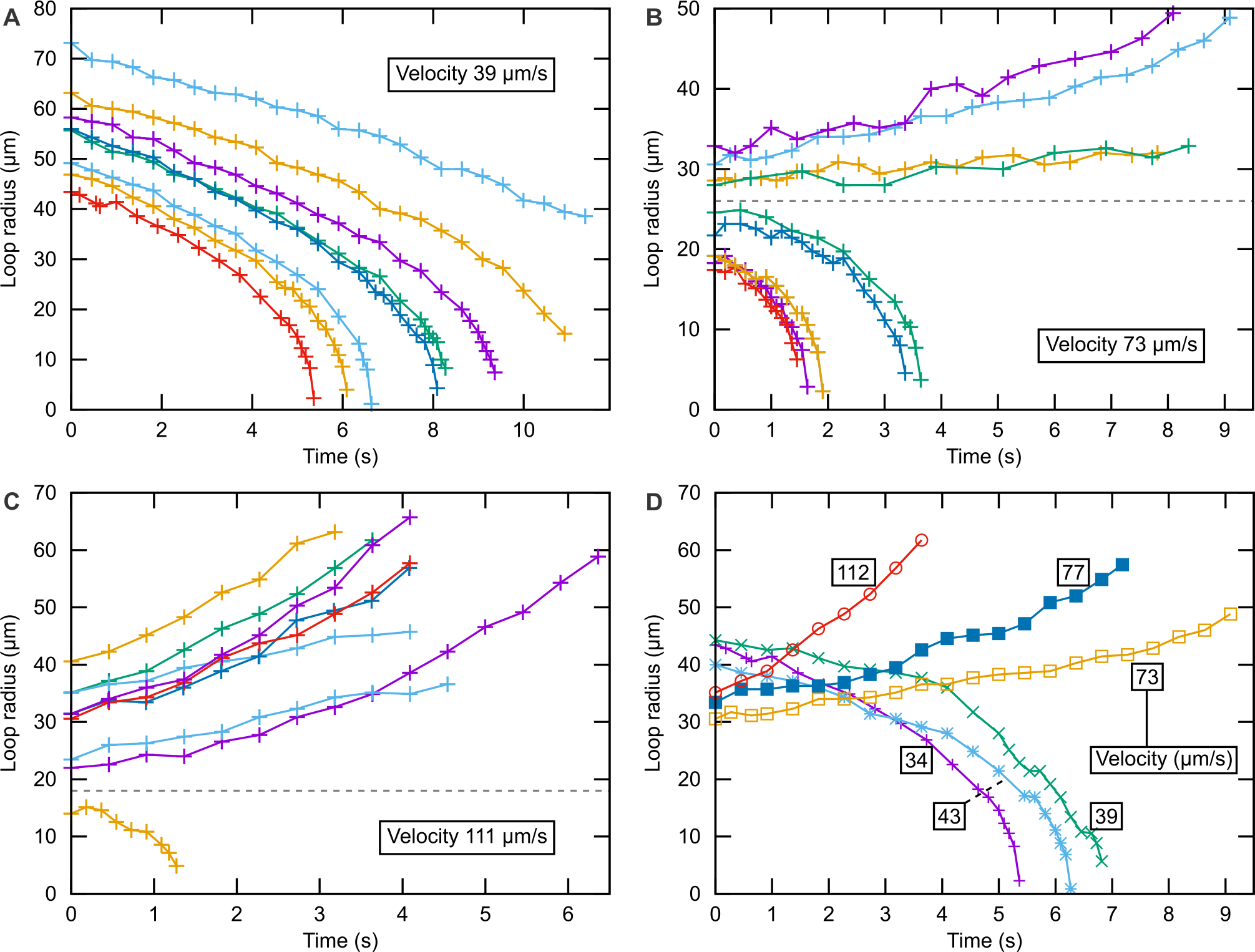}
\caption{\small
{\bf An independent study, performed in $400~{\rm \mu m}$ wide and $15~{\rm \mu m}$ deep channels, shows equivalent behaviour of dowser domain dynamics as is discussed in the main text.}
({\bf A-C}) Time dependence of the domain radius for three different flow velocities. Dowser domains with the initial radius above the critical size (approximate value is shown by the dashed line) grow, whereas domains below the critical size shrink.
({\bf D}) At similar initial sizes, domains in flows with low velocity shrink, while domains at large flow velocity grow in time. Solid lines are added to guide the eye.
}
\end{figure*}

\begin{figure*}[!htb]
\centering \includegraphics[width=152mm]{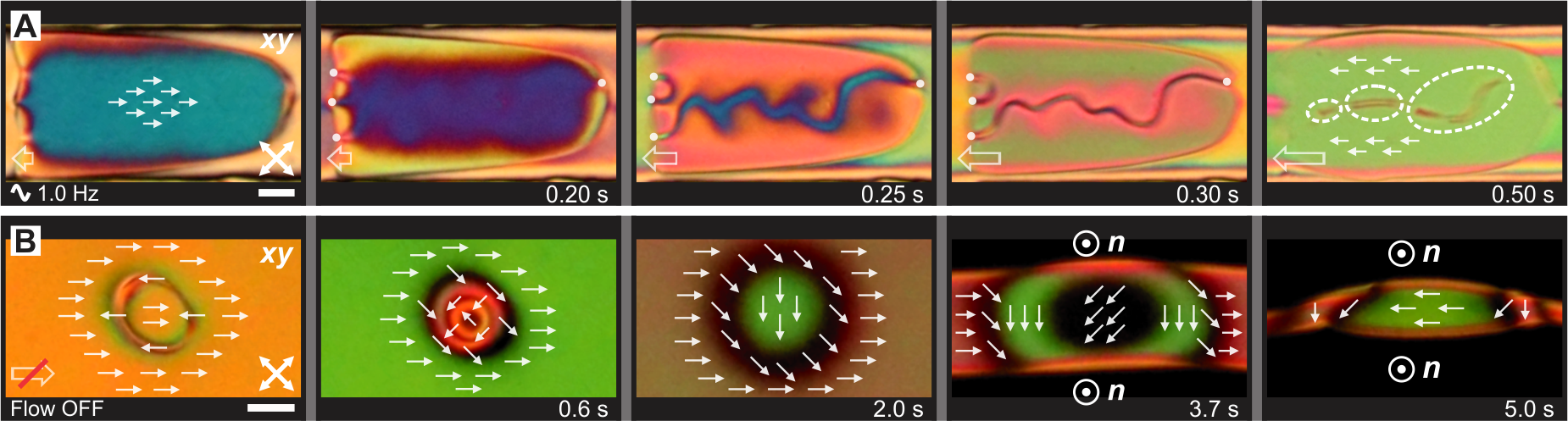}
\caption{\small
{\bf Dowser field relaxation dynamics.}
({\bf A}) For sufficiently rapid flow rate changes, the director reversal behaves as a quench. It is instantaneous across the entire domain and leads to a nucleation of irregularly shaped solitons, which coarsen over time. While the solitons are topologically protected, they disappear when the characteristic length is so small that a discontinuous director rearrangement becomes possible (dashed ellipses). This scenario effectively represents a local breakdown or ``melting'' of the dowser alignment, which happens in a fraction of a second.
({\bf B}) A closed circular soliton unwinds and enlarges in a process akin to diffusion of orientational order after shutting off the flow, as Eq.~(1) reduces to a diffusion equation in the absence of flow. A slight transient increase of the winding angle is observed in the fourth panel, due to a rebound of the liquid flow, which makes the leftward orientation preferable, as also suggested by the bulbous shape in the final panel. Scale bars, $20~{\rm \mu m}$.
}
\end{figure*}
\end{small}
\clearpage

\end{document}